# Minimum Vertex-type Sequence Indexing for Clusters on Square Lattice

Longguang Liao[1], Yujun Zhao[1,2], Zexian Cao[3], and Xiaobao Yang[1,2]★

An effective indexing scheme for clusters that enables fast structure comparison and congruence check is desperately desirable in the field of mathematics, artificial intelligence, materials science, etc. Here we introduce the concept of minimum vertex-type sequence for the indexing of clusters on square lattice, which contains a series of integers each labeling the vertex type of an atom. The minimum vertex-type sequence is orientation independent, and it builds a one-to-one correspondence with the cluster. By using minimum vertex-type sequence for structural comparison and congruence check, only one type of data is involved, and the largest amount of data to be compared is *n* pairs, *n* is the cluster size. In comparison with traditional coordinate-based methods and distance-matrix methods, the minimum vertex-type sequence indexing scheme has many other remarkable advantages. Furthermore, this indexing scheme can be easily generalized to clusters on other high-symmetry lattices. Our work can facilitate cluster indexing and searching in various situations, it may inspire the search of other practical indexing schemes for handling clusters of large sizes.

[1] Department of Physics, South China University of Technology, Guangzhou 510640, China.

[2] Key Laboratory of Advanced Energy Storage Materials of Guangdong Province, South China University of Technology, Guangzhou 510640, China. [3] Institute of Physics, Chinese Academy of Sciences, Beijing 100190, China. ★ e-mail: scxbyang@scut.edu.cn

Clusters containing a few to several thousands of atoms serve to bridge the gap between an isolated atom and the bulk counterpart[1-3]. They may exhibit peculiar physical and chemical properties that are probably not to be observed in the bulk materials. Small clusters are anticipated to exhibit strongly size-dependent features, while those intermediate- and large-sized clusters might exhibit a smoothly varying behavior which approaches the bulk limit[3-5]. The size and structure are the fundamental factors in determining the properties of a cluster, they are thus of great concern in the study and application of clusters.

The conventional experimental methods for structural determination are hardly able to directly obtain the atomic configuration of clusters[6-10], as collaborating with theoretical computation is usually a necessity[1-3, 11], yet without a guarantee of success. This is of no surprise since finding the global minimum on a given potential-energy-surface, a key criterion for predicting ground state structure of a cluster, is essentially a formidable task. The number of local minima increases exponentially with the increasing cluster size[3, 12, 13], as it is obviously an NP-hard problem that can be mapped to the traveling-salesman problem[12, 14]. In the past two decades, several global optimization methods of different efficiencies for cluster structure prediction have been devised, including simulated annealing[15], basin/minima hopping[16-18], genetic algorithm[18-26], particle swarm optimization algorithm[27-32], etc. To investigate the structural evolution by global optimization, a large number of trial structures are to be generated at each generation. Consequently, it is highly desirable to develop some techniques to label the transient structures and to judge the similarity or congruency among the structures available at the intermediate stages, which serve to avoid futile, repetitious computation so as to effectively accelerate the search process[32]. A complete sampling of all the minima on a potential-energy-surface is simply impossible, but a high-throughput screening under restricted conditions can be still very helpful[13, 33], for which the congruence check of clusters is a crucial prerequisite. In order to congruence check two distinct clusters, it needs to establish a one-to-one mapping, or correspondence, between their structural elements. The most straightforward way would be to compare the coordinates of the corresponding atoms in those two clusters[34]. This coordinate-based method is very complicated because both clusters are subjected to a translation operation so that their centroids are at the origin of the coordinate system, and one of them needs an additional rotation to minimize the deviation in the corresponding coordinates. For two 2D clusters each containing $n$ atoms, the largest amount of comparing data this way is significantly greater than $2n$ pairs (for square lattice, without distinguishing the enantiomers, it is usually $16n$ pairs). An alternative strategy is to compare the corresponding interatomic distances in the two clusters, i.e., the distance-matrix method[29, 34]. However, as Lv et al.[32] once pointed out, "the distance metric requires ordering of atoms in a structure, and thus is not able to unambiguously fingerprint structures." For two clusters of size $n$, the largest amount of comparing data in this way is much more than $n(n-1)/2$ pairs. Other methods such as those involving Bond-Characterization-Matrix[31, 32], Zernike descriptors[35] and spherical harmonic descriptors[36], can provide an approximate, quantitative measure for structural similarity between two clusters, but cannot label the distinct structures. The aforementioned methods are all suitable for both amorphous and crystalline clusters. For 2D and 3D high-symmetry crystal-fragment clusters, where the interatomic distances among the nearest neighboring atoms are fixed, there is also another recognition technique based on the relative orientations of the bonds, which employs codes obtained by using the Balaban and von Schleyer's technique[37, 38]. In this approach the largest amount of comparing data is reduced to $(n-1)$ pairs, it is, however, not a fast structure retrieval method since finding the main and side chains of a given structure is very time-consuming.

In the current work we introduce a new indexing scheme, the minimum vertex-type sequence, to label and characterize the clusters on square lattice (below often simply referred to as cluster when no ambiguity may arise), which might meet some requirements for fast and effective decision making of artificial intelligence as in the game of Go[39]. This indexing scheme employs only the vertex type, or precisely the nearest-neighbor configuration, for each atom in a cluster. The atoms in a cluster are ordered and labeled following a special rule that the cluster can be indexed with a unique digit sequence, being independent from either the choice of the reference frame or the orientation of a cluster in given configuration. A one-to-one correspondence between the minimum vertex-type sequence and the cluster can be established. For clusters of size $n$, the largest amount of comparing data for congruency check is only $n$ pairs.

**Results**

Below we will demonstrate that for crystal-fragment clusters, the vertex type sequence, which characterizes the nearest-neighbor configuration of each atom in a cluster, provides a practicable algorithm for fast structure comparison. The minimum vertex-type sequence method will be demonstrated in detail by treating clusters on square lattice.

Firstly, we classify the vertexes for atoms in a cluster on square lattice based on their nearest-neighbor configuration (the orientations of the bonds with the nearest-neighbors are discriminated). There are only fifteen different vertex types for clusters on square lattice, which are labeled accordingly with the integers 1-15 in Figure 1.

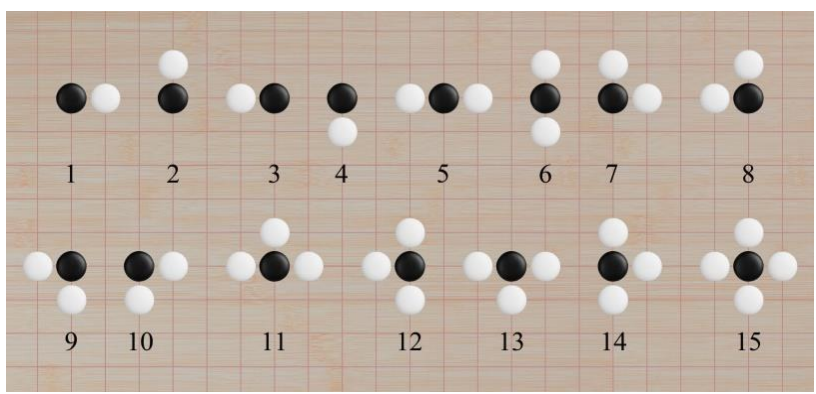

**Figure 1** | The fifteen possible vertex types appearing in a cluster on square lattice, as labeled by integers 1-15. The black stones denote the atoms of concern, while the white stones indicate the existence of possible nearest neighbors.

Secondly, atoms in the cluster are labeled with their vertex types in the order of

"left-to-right-and-bottom-up", i.e., the atom at the bottom left corner is set to be the first atom, it is then the turn of atoms in the same horizontal line from left to right till the end of that line, which is to be continued from the leftmost atom of the next upper line (if there is any). This process will proceed to the atom at the upper right corner of the cluster (Fig.2).

By assigning to all the atoms a corresponding vertex type in the aforementioned order, we obtain an $n$-digit vertex-type sequence, where $n$ is the size of the cluster. Since the vertex types are discriminated with regard to the orientations of bonds to the nearest-neighbors, the same cluster in different orientations may have distinct vertex-type sequences. As the space group for square lattice is group $D_4$ with eight elements, a cluster here concerned may have at most eight different vertex-type sequences. In Fig.2 displayed are the different configurations of a cluster of 4 atoms under the action of group $D_4$, the eight vertex-type sequences are (1,8,6,4), (2,1,5,9), (2,6,10,3), (7,5,3,4), (7,3,6,4), (1,5,8,4), (2,6,1,9) and (2,10,5,3), respectively.

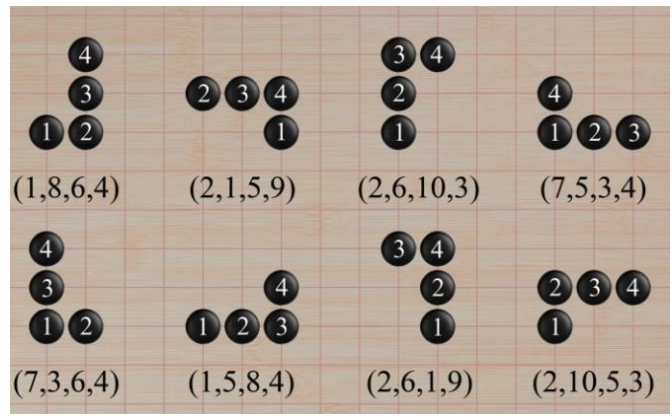

**Figure 2** | The different orientations of a given configuration for 4-atom cluster under the action of group $D_4$ for square lattice and the corresponding vertex-type sequences. The minimum vertex-type sequence is (1,5,8,4).

Among those eight (or less if the clusters possess rotation and/or reflection symmetries) vertex-type sequences, we adopt the one which has the minimum preceding numbers as the character of the cluster, and denote it as the minimum vertex-type sequence. For the cluster of 4 atoms shown in Fig.2, the minimum number for the first digit is 1, and in the two sequences beginning with 1, i.e., (1,8,6,4) and

(1,5,8,4), the minimum number for the second digit is 5, thus the minimum vertex-type sequence for this cluster is (1,5,8,4).

It can be strictly demonstrated that there is a one-to-one mapping between the minimum vertex-type sequence and the cluster (see the supplementary information for details). This is to say that the minimum vertex-type sequence can be used to characterize the individual clusters on square lattice, which can serve the identification and comparison of the clusters. With the minimum vertex-type sequence, the largest amount of data involved in the comparison of two clusters of $n$ atoms can be reduced to only $n$ pairs.

Table 1. Comparison of the four structural description schemes for clusters on high-symmetry lattice. CB: coordinate-based method; DM: distance-matrix method; BS: Balaban & von Sheleyer's technique; MVTS: minimum vertex-type sequence.

| Method | Information | Orientation independence | Labeling of atoms | Number of data types | Complexity of data sampling | Data size |
|--------|-------------|--------------------------|-------------------|---------------------|------------------------------|-----------|
| CB     | coordinates | No                       | N/A               | 1                   | easy                         | $mn$[a]   |
| DM     | distances   | Yes                      | Ambiguous         | 1                   | easy                         | $(n^2-n)/2$ |
| BS     | Bond orientations | Yes                | definite          | 3                   | hard                         | $n-1$[b]  |
| MVTS   | Vertex types | Yes                     | definite          | 1                   | easy                         | $n$       |

[a] $m$ denotes the dimension of the system concerned;

[b] If a structure is formed only from the main chain, the size of the descriptive data is ($n$-1); if a structure is branched, the data size is then greater than (n-1).

For the comparison of two clusters, this minimum vertex-type sequence scheme has two distinct advantages over the conventional coordinate-based methods. It uses only a few descriptive data and, more importantly, it is orientation independent. Unlike the distance-matrix method which suffers from ambiguous ordering of the atoms[32], the labeling of atoms in the minimum vertex-type sequence is concise and determinate. In comparison with the Balaban and von Sheleyer's technique[37, 38], our minimum vertex-type sequence scheme employs only one single type of data, and it avoids the difficult process of finding out the main chains of a cluster, which is usually a formidable task. Roughly speaking, our minimum vertex-type sequence scheme uses only a series of integers labeling the vertex type of atoms, whereas the Balaban and von Sheleyer's technique employs three types of data, i.e., digits denoting the orientation of the bonds, parentheses denoting the branches and commas denoting the different branches on the same level. An overview of the comparison of the minimum vertex-type sequence scheme to those conventional techniques is summarized in Table 1. Obviously, the minimum vertex-type sequence scheme can serve the fast congruence-check for clusters on high-symmetry lattices.

**Discussion**

The minimum vertex-type sequences build a one-to-one correspondence with clusters on square lattice, thus a minimum vertex-type sequence can be regarded as the index tag of the corresponding cluster. That is to say that each cluster can be characterized by a unique minimum vertex-type sequence. For example, the clusters of size 2 can be indexed as (1, 3). The cluster of size 4 displayed in Fig.2, which is one of the five possible configurations, can be indexed as (1,5,8,4). For clusters of size 7 in the two configurations in Fig.3a (there are 108 different configurations in total), the minimum vertex-type sequences are (1,5,8,2,10,5,9) and (1,5,8,6,1,5,9), respectively. Comparing with the Balaban and von Schleyer's indexing system[37, 38], our minimum vertex-type sequence indexing system has at least two advantages. Firstly, the process of obtaining the minimum vertex-type sequence of a cluster is simpler than that to get the Balaban and von Schleyer's codes which usually require the determination of the main and the side chain(s) in a branched cluster. Secondly, the minimum vertex-type sequence indexing system is more economical for computer calculation, since the minimum vertex-type sequences use only one type of data, whereas the Balaban and von Schleyer's codes usually use three types of data.

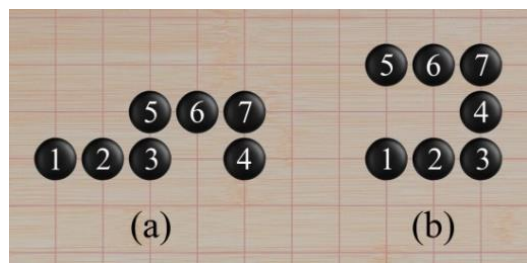

**Figure 3 |** The two configurations for cluster of 7 atoms corresponding to the minimum vertex-type sequence (1,5,8,2,10,5,9) (a), and (1,5,8,6,1,5,9) (b), respectively.

In a practical implementation of congruence check for clusters of large sizes, approximate representation of single-number indexing can be used as a rapid primary filter, which is to be further confirmed by comparing the minimum vertex-type sequences. A good choice for the approximate representation of single-number indexing is to let $f(P) = \sum_{i=1}^{n} S_i \cdot i^6$, where $S_i$ is the $i$-th digit in the minimum vertex-type sequence of the cluster P. The final decision can be made by digit-by-digit comparing the corresponding minimum vertex-type sequences among the chosen candidates.

The minimum vertex-type sequence indexing scheme is orientation independent and employs only $n$ integers to describe the configuration of a cluster on the square lattice. It is anticipated to provide a fast structure recognition method for large clusters. The excellent effectiveness of this indexing scheme can be verified in solving this typical problem: finding out all the isomers on square lattice for clusters as large as possible in a tolerable running time. This is a formidable task because the number of isomers increases exponentially with the cluster size[3, 12, 13]. When the cluster size reaches 13, as derived with the current method, there are 238591 types of isomers (with the enantiomers undistinguished). Each time when a new structure is generated, it needs to be compared with a great amount of existing isomers to judge whether a new type has occurred or not, which might immediately use up the computer resource. Our minimum vertex-type sequence indexing scheme presented above provides a preferable fast structure retrieval method.

The minimum vertex-type sequence indexing also has many other merits. A minimum vertex-type sequence contains the information of vertex types of atoms, i.e., the nearest-neighbor configurations of atoms. It can be easily converted into the information of bonds. Therefore, the minimum vertex-type sequence indexing is preferable in the situation where the bond energy[4] is of concern.

Furthermore, the concept of minimum vertex-type sequence can be generalized to other high-symmetry lattices such as the regular triangular lattice, the honeycomb lattice, the cubic lattice, and the diamond lattice. In the case of square lattice there are 15 possible vertex types to be handled, while for regular triangular lattice there are 63, and for honeycomb lattice there are 14 (more details will be presented in a forthcoming publication). By combining the minimum vertex-type sequence method with high-throughput screening procedure under restricted conditions, effective strategies can be devised to search new atomic cluster structures on those complicated lattices.

We devised the minimum vertex-type sequence indexing scheme to characterize clusters on square lattice. The minimum vertex-type sequence of a cluster comprises only one integer for each atom to label its vertex type, and the sequence is orientation independent. It turns out to be the most preferable scheme for fast recognition and strict comparison of clusters, as it employs fewer and simpler descriptive data. In particular, for fast preliminary filtering, the minimum vertex-type sequences can be at first converted into a single integer, which can further accelerate the procedure of structure comparison dramatically. Our work is anticipated to initiate the search of other practical indexing schemes for handling clusters of large sizes, which are desperately desired in many fields.

## Acknowledgments

We thank Prof. Jiang Nan for his kind and very professional help in preparing the figures. This work was supported by NSFC grant nos. 11474100, 11474335, and 11574088; by the Guangdong Natural Science Funds for distinguished young scholars grant no. 2014A030306024, and by the Fundamental Research Funds for central universities grant no. 2015PT017. The national supercomputing center at Shenzhen and the super-computing center of Chinese Academy of Sciences are gratefully acknowledged.

**Author Contributions**

L.G. L., Y.J.Z. and X. B. Y. designed the research. L.G.L. did the computation. L.G.L and Z.X.C. discussed the results and compiled the manuscript.

**Competing financial interests**

The authors declare no competing financial interests

**References**

1. Brack, M. The physics of simple metal clusters: self-consistent jellium model and semiclassical approaches. *Rev. Mod. Phys.* **65**, 677-732 (1993).
2. de Heer, W.A. The physics of simple metal clusters: experimental aspects and simple models. *Rev. Mod. Phys.* **65**, 611-676 (1993).
3. Baletto, F. & Ferrando, R. Structural properties of nanoclusters: Energetic, thermodynamic, and kinetic effects. *Rev. Mod. Phys.* **77**, 371-423 (2005).
4. Yang, X., Zhao, Y.-J., Xu, H. & Yakobson, B.I. Ground states of group-IV nanostructures: Magic structures of diamond and silicon nanocrystals. *Phys. Rev. B* **83**, 205314 (2011).
5. Li, S.F., Zhao, X.J., Xu, X.S., Gao, Y.F. & Zhang, Z. Stacking Principle and Magic Sizes of Transition Metal Nanoclusters Based on Generalized Wulff Construction. *Phys. Rev. Lett.* **111**, 115501 (2013).
6. Wolf, M.D. & Landman, U. Genetic Algorithms for Structural Cluster Optimization. *J. Phys. Chem. A* **102**, 6129-6137 (1998).
7. Doye, J.P.K. Identifying structural patterns in disordered metal clusters. *Phys. Rev. B* **68**, 195418 (2003).
8. Woodley, S.M. & Catlow, R. Crystal structure prediction from first principles. *Nature Mater.* **7**, 937-946 (2008).
9. Catlow, C.R.A. *et al.* Modelling nano-clusters and nucleation. *Phys. Chem. Chem. Phys.* **12**, 786-811 (2010).
10. Woodley, S.M., Hamad, S. & Catlow, C.R.A. Exploration of multiple energy landscapes for zirconia nanoclusters. *Phys. Chem. Chem. Phys.* **12**, 8454-8465 (2010).
11. Marks, L.D. Experimental studies of small particle structures. *Rep. Prog. Phys.* **57**, 603 (1994).
12. Wille, L.T. & Vennik, J. Computational complexity of the ground-state determination of atomic clusters. *J. Phys. A: Math. Gen.* **18**, L419 (1985).
13. Xu, S.-G., Zhao, Y.-J., Liao, J.-H. & Yang, X.-B. Understanding the stable boron clusters: A bond model and first-principles calculations based on high-throughput screening. *J. Chem. Phys.* **142**, 214307 (2015).
14. Papadimitriou, C.H. The Euclidean travelling salesman problem is NP-complete. *Theor. Comp. Sci.*

**4**, 237-244 (1977).

15. Kirkpatrick, S., Gelatt, C.D. & Vecchi, M.P. Optimization by Simulated Annealing. *Science* **220**, 671-680 (1983).
16. Wales, D.J. & Doye, J.P.K. Global Optimization by Basin-Hopping and the Lowest Energy Structures of Lennard-Jones Clusters Containing up to 110 Atoms. *J. Phys. Chem. A* **101**, 5111-5116 (1997).
17. Goedecker, S. Minima hopping: An efficient search method for the global minimum of the potential energy surface of complex molecular systems. *J. Chem. Phys.* **120**, 9911-9917 (2004).
18. Schönborn, S.E., Goedecker, S., Roy, S. & Oganov, A.R. The performance of minima hopping and evolutionary algorithms for cluster structure prediction. *J. Chem. Phys.* **130**, 144108 (2009).
19. Hartke, B. Global geometry optimization of clusters using genetic algorithms. *J. Phys. Chem.* **97**, 9973-9976 (1993).
20. Deaven, D.M. & Ho, K.M. Molecular Geometry Optimization with a Genetic Algorithm. *Phys. Rev. Lett.* **75**, 288-291 (1995).
21. Pullan, W.J. Genetic operators for the atomic cluster problem. *Comput. Phys. Commun.* **107**, 137-148 (1997).
22. Hartke, B. Global cluster geometry optimization by a phenotype algorithm with Niches: Location of elusive minima, and low-order scaling with cluster size. *J. Comput. Chem.* **20**, 1752-1759 (1999).
23. Chen, Z., Jiang, X., Li, J. & Li, S. A sphere-cut-splice crossover for the evolution of cluster structures. *J. Chem. Phys.* **138**, 214303 (2013).
24. Oganov, A.R. & Glass, C.W. Crystal structure prediction using ab initio evolutionary techniques: Principles and applications. *J. Chem. Phys.* **124**, 244704 (2006).
25. Glass, C.W., Oganov, A.R. & Hansen, N. USPEX—Evolutionary crystal structure prediction. *Comput. Phys. Commun.* **175**, 713-720 (2006).
26. Lyakhov, A.O., Oganov, A.R., Stokes, H.T. & Zhu, Q. New developments in evolutionary structure prediction algorithm USPEX. *Comput. Phys. Commun.* **184**, 1172-1182 (2013).
27. Eberhart, R. & Kennedy, J. A new optimizer using particle swarm theory. *Proceedings of the Sixth International Symposium on Micro Machine and Human Science, MHS '95*, 39-43 (1995).
28. Kennedy, J. & Eberhart, R. Particle swarm optimization. *IEEE International Conference on Neural Networks* **4**, 1942-1948 (1995).
29. Call, S.T., Zubarev, D.Y. & Boldyrev, A.I. Global minimum structure searches via particle swarm optimization. *J. Comput. Chem.* **28**, 1177-1186 (2007).
30. Wang, Y., Lv, J., Zhu, L. & Ma, Y. Crystal structure prediction via particle-swarm optimization. *Phys. Rev. B* **82**, 094116 (2010).
31. Wang, Y., Lv, J., Zhu, L. & Ma, Y. CALYPSO: A method for crystal structure prediction. *Comput. Phys. Commun.* **183**, 2063-2070 (2012).
32. Lv, J., Wang, Y., Zhu, L. & Ma, Y. Particle-swarm structure prediction on clusters. *J. Chem. Phys.* **137**, 084104 (2012).
33. Curtarolo, S. *et al.* The high-throughput highway to computational materials design. *Nature Mater.* **12**, 191-201 (2013).
34. Maiorov, V.N. & Crippen, G.M. Significance of root-mean-square deviation in comparing three-dimensional structures of globular proteins. *J. Mol. Biol.* **235**, 625-634 (1994).
35. Kihara, D., Sael, L., Chikhi, R. & Esquivel-Rodriguez, J. Molecular surface representation using

3D Zernike descriptors for protein shape comparison and docking. *Curr. Protein Peptide Sci.* **12**, 520-530 (2011).

36. Funkhouser, T. *et al.* A search engine for 3D models. *ACM Trans. Graph.* **22**, 83-105 (2003).

37. Balaban, A.T. & von R. Schleyer, P. Systematic classification and nomenclature of diamond hydrocarbons—I: Graph-theoretical enumeration of polymantanes. *Tetrahedron* **34**, 3599-3609 (1978).

38. Filik, J. Diamondoid Hydrocarbons, in: Carbon based nanomaterials. Materials science foundations (monograph series), Vol. 65-66, 1-26. (eds. N. Ali, A. Ochsner & W. Ahmed) (Trans Tech, Switzerland; 2010).

39. Silver, D. *et al.* Mastering the game of Go with deep neural networks and tree search. *Nature* **529**, 484-489 (2016).

# Supplementary Information

## Minimum Vertex-type Sequence Indexing for Clusters on Square Lattice

Long-Guang Liao,[1] Yu-Jun Zhao,[1,2] Ze-Xian Cao,[3] and Xiao-Bao Yang[1,2]★

This supplementary information includes the demonstration of the one-to-one correspondence between a cluster on square lattice and its minimum vertex-type sequence, with illustrating examples. In the main text it has been shown that a cluster can be converted into one and only one minimum vertex-type sequence. Here we need only to show that for a given minimum vertex-type sequence, one and only one cluster can be constructed thereupon.

The first item of business for us is to decompose a given minimum vertex-type sequence, i.e., to remove the digits one by one from the sequence till the initial digit of the original sequence. This procedure corresponds to the process of stripping off atoms from a cluster one by one till the original atom. Through this procedure, a series of daughter sequences with ever decreasing number of digits, and some of the remaining digits in the sequence have to be accordingly adjusted (see below), can be obtained. If it is always the last digit that is to be removed at each step, the thus derived series of daughter sequences is unique.

The second item of business is to construct the cluster by adding atoms one by one, based on the series of daughter sequences obtained in the foregoing decomposition process. It will turn out that the procedure of construction is also unique.

**Decomposing a minimum vertex-type sequence**

Before starting the decomposition procedure for a minimum vertex-type sequence, let's first analyze the possible vertex types for the last atom of any clusters on square lattice.

It is easy to see that according to the "left-to-right-and-bottom-up" ordering rule, the possible vertex type for the last atom of a cluster on square lattice can be only one among the choices of 3, 4 and 9 (see Fig.1). This is to say that all the minimum vertex-type sequences must end with a digit of 3, 4 or 9 (Cf. Fig.2).

In the case (1) that the last atom P in the cluster is of vertex type 3, this atom P has only one nearest neighbor, the atom O, sitting directly left to atom P. Since there must be one atom sitting directly right to but without any atom sitting directly above it, the possible vertex types for the atom O can be only one among 1, 5, 10 and 13 (see Fig.1). Consequently, for a minimum vertex-type sequence ending with 3, the penultimate digit must be one among 1, 5, 10 and 13. After stripping off the last atom P of vertex type 3 from the cluster, the vertex type of the penultimate atom O will undergo the subsequent change: 5→3; 10→4; 13→9; and particularly 1→0. Here the auxiliary vertex type 0 refers to an isolated atom. An isolated atom, of vertex type 0, can appear in the residual fragments after stripping off atoms from a cluster of finite size.

In the case (2) that the last atom P is of vertex type 4, it has only one nearest neighbor atom, the atom Q, sitting directly below it. Having one atom directly above it, atom Q can assume one of the vertex types 2, 6, 7, 8, 11, 12, 14 or 15. It needs to point out here that among those atoms having a nearest neighbor directly above it, the atom Q is the last one in the cluster. As a result, in a minimum vertex-type sequence with the last digit being 4, there must be at least one digit being among the eight digits 2, 6, 7, 8, 11, 12, 14, and 15. Thus the last digit in the sequence being one among 2, 6, 7, 8, 11, 12, 14 and 15 refers to the atom Q in the cluster. After stripping off atom P of vertex type 4 from a cluster, the vertex type for the penultimate atom Q will consequently change this way: 6→4; 7→1; 8→3; 11→5; 12→9; 14→10; 15→13; and in particular 2→0.

In the case (3) that the last atom P is of vertex type 9, atom P has two nearest neighbors: One is the penultimate atom O sitting directly left to atom P, the other is atom Q sitting directly below atom P. The possible vertex types for atom O, as well as their transmutation after the stripping off of atom P, are just the same as in case (1) where the last atom P is of vertex type 3. The possible vertex types for atom Q, as well as their transmutation after the stripping off atom P, are just the same as in case (2) where the last atom P is of type 4. Note that the atom Q is the very last one among those atoms that may have a nearest neighbor sitting directly above it. Thus, for a minimum vertex-type sequence with the last digit being 9, the penultimate digit must be one among 1, 5, 10 and 13, and it must also contain at least one digit among the eight possibilities of 2, 6, 7, 8, 11, 12, 14 and 15. Any one of these eight possibilities that appears the latest in a minimum vertex-type sequence corresponds to atom Q.

In addition to the above three cases referring to vertex types 3, 4 and 9, there is another possibility for the last digit which is the ancillary vertex type 0 denoting an isolated atom. In this case, stripping off the atom referred to vertex type 0 does not affect the vertex type of any other atoms. Or we can say that by removing the last digit being 0 from a daughter sequence none of the other digits will be altered. Note that when there is only one digit left in a daughter sequence, this digit can be only 0.

Following the aforementioned digit-stripping rules, we can now decompose a given minimum vertex-type sequence by removing the digits one by one. Writing down the derived daughter sequence in each step, which has been partially altered with regard to the corresponding piece in the original minimum vertex-type sequence, we thus obtain a series of daughter sequences each with one less digit. For example, the minimum vertex-type sequence derived from the cluster shown in Fig.3a can be decomposed this way: (1,5,8,2,10,5,9) → (1,5,8,0,10,3) → (1,5,8,0,4) → (1,5,3,0) → (1,5,3) → (1,3) → (0).

From the above discussion, we see that the process of decomposing a minimum vertex-type sequence is unique, since the derived series of daughter sequences refers to the process of stripping off atoms one by one from the cluster, and the series of the residual cluster fragments is unique.

**Constructing cluster from a minimum vertex-type sequence**

A unique cluster can be constructed based on the series of daughter sequences of a minimum vertex-type sequence. The correspondence between the n-digit minimum vertex-type sequence and the atom configuration in a cluster of size n is guaranteed, but not readily to be recognized. Fortunately, the decomposition of the minimum vertex-type sequence into the series of daughter sequences provides a step-by-step guidance for the construction of the cluster from scratch.

It begins with an isolated atom, of vertex type 0. This atom 1 can be placed on any lattice point on the square lattice. Without loss of generality, suppose it is put on the origin of the coordinate system for the square lattice. Then more atoms will be added one by one to the acquired structure. Below we will show that the relative position of the atom to be added at each stage can be determined by two successive daughter sequences previously obtained by decomposing a given minimum vertex-type sequence.

Let's consider the addition of the k-th atom to an existing cluster fragment of $(k-1)$ atoms, where k>1 is a positive integer. Four different situations may be confronted by this atom $k$.

In the case that the last digit is 3 in the k-digit daughter sequence, the atom $k$ should be placed on the right side of the $(k-1)$-th atom. In this case, the k-digit daughter sequence differs from the (k-1)-digit daughter sequence only in its (k-1)-th digit, of course it now also has the additional k-th digit, as illustrated by panels a-b and panels b-c of Fig.S1, respectively.

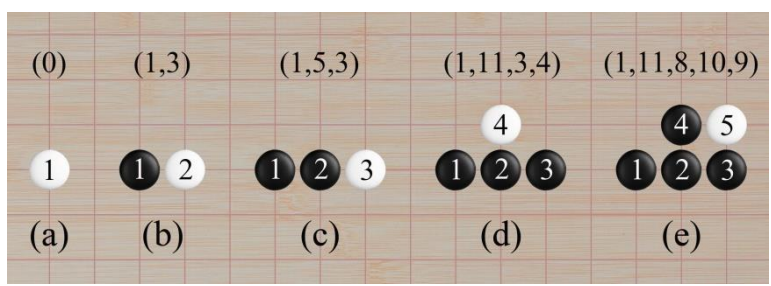

**Supplementary Figure 1**. Constructing a cluster according to the series of daughter sequences of a minimum vertex-type sequence. The white stones indicate the newly added atoms at each step.

In the case that the last digit of the k-digit daughter sequence is 4, the atom $k$ should be placed directly above some $g$-th atom ($g < k$). To determine the value of g, we notice that the $k$-digit daughter sequence differs from the preceding $(k-1)$-digit daughter sequence only at the $g$-th digit, and has the additional $k$-th digit. For example, given that the successive daughter sequences of 3- and 4 digits be (1,5,3) and (1,11,3,4), respectively, we know that atom 4 should be placed directly above atom 2, as illustrated in panels c-d of Fig.S1.

In the case that the k-digit daughter sequence terminates with 9, the atom k should be placed on the right-hand-side of the $(k-1)$-th atom. At the same time, it also should be directly above some $g$-th atom ($g < k-1$). To determine the value of g, we notice that the $k$-digit daughter sequence differs from $(k-1)$-digit daughter sequence at the $g$-th and $(k-1)$-th digits, and has the additional $k$-th digit. For example, given that the two successive sequences of 4 and 5 digits be (1,11,3,4) and (1,11,8,10,9), then atom 5 should be placed on the right hand side of atom 4 and meanwhile directly above atom 3, as shown in panels d-e of Fig.S1.

In the particular case that the last digit of the $k$-digit daughter sequence is 0, the relative position of atom $k$ with respect to the atoms in the existing fragment cannot be immediately determined, yet the problem can be finally settled on the basis of the whole series of daughter sequences because of the integrity of a cluster. To determine the position of the atom $k$, information from a few lengthy daughter sequences is demanded. In practice, we can proceed along the following line. The first atom of vertex type 0, i.e. atom 1, is placed at the origin. Atoms having a definite position relative to atom 1, including this atom 1, are grouped into branch one. The second atom of vertex type 0, labeled as atom B for the moment, will be temporarily placed at a point that is sufficiently distant to avoid any overlap between branch one and the forthcoming branch to grow around atom B. The forthcoming atoms having a definite position with regard to atom B are grouped into branch two. Other isolated atoms of vertex type 0, if there are any, and the forthcoming atoms, are handled in the same way. When a freshly added atom connects two existing distinct branches, the atoms in the younger branch should be taken to merge into the elder one. At the final stage of construction, the cluster as an integral unity should have just one branch.

For example, let's construct a cluster from a given series of daughter sequences of (0) → (1,3) → (1,5,3) → (1,5,8,4) → (1,5,8,4,0) → (1,5,8,4,1,3) → (1,5,8,6,1,5,9) (for the decomposition procedure the series is presented in the reversed order). It proceeds as follows: Atom 1 is placed on the origin; atom 2 is placed directly right to atom 1; atom 3 is placed directly right to atom 2; and atom 4 is placed directly above atom 3. These four atoms have definite relative positions with respect to atom 1, and they are grouped into branch one having a size of 4. The last digit of the 5-digit daughter sequence is 0. Thus, atom 5 is the second atom of vertex type 0, it is temporarily placed at a faraway point to guarantee non-overlapping with branch one. Atom 6 is placed directly right to atom 5. So far, atoms 5 and 6 are grouped into branch two. Next we see that the atom 7 should sit directly above atom 4, and meanwhile right to atom 6—therefore the atom 7 connects the two branches. The thus obtained cluster is shown in Fig.3b.

From the above analysis we see that the relative position of each atom with respect to atom 1 can be determined on the basis of the series of daughter sequences obtained through the decomposition of a given minimum vertex-type sequence. It is definite and unique.

A true minimum vertex-type sequence for clusters on square lattice needs meet two requirements: (1) It can be decomposed into a series of daughter sequences each with one less digit; (2) A one-piece cluster can be constructed following the rules of construction on the basis of the series of daughter sequences. The one-piece cluster can be encoded into a minimum vertex-type sequence which reproduces the original series of daughter sequences. For such a minimum vertex-type sequence, the procedure for both decomposition and construction is unique.